\documentclass{elsart3p}

\usepackage{graphicx}
\usepackage{epsfig}

\usepackage{amssymb}

\begin{document}

\begin{frontmatter}

% 7 pages, 3 figures, accepted for publication in Phys. Lett. A.

%titles, authors and addresses

% \address{Address\thank

\title{Pressure dependence of superconductivity in doped two-leg
ladder cuprates}


\author{Jihong Qin}

\address{Department of Physics, Beijing University of Science and
Technology, Beijing 100083, China}

\author{Ting Chen and Shiping Feng}
\address{Department of Physics, Beijing Normal University, Beijing
100875, China}

\maketitle
%\date{\today}
\begin{abstract}
Within the kinetic energy driven superconducting mechanism, the
effect of the pressure on superconductivity in the doped two-leg
ladder cuprates is studied. It is shown that the superconducting
transition temperature in the doped two-leg ladder cuprate
superconductors increases with increasing pressure in the
underpressure regime, and reaches a maximum in the optimal
pressure, then decreases in the overpressure regime. This domed
shape of the pressure dependence of the superconducting transition
temperature is similar to that of the pressure dependence of the
longitudinal part of the superconducting gap parameter, indicating
that the pressure dependence of superconductivity in the doped
two-leg ladder cuprates is mainly produced by the development of
the pairing correlation along legs.
\end{abstract}

\end{frontmatter}

After over ten years of intense experimental study of the doped
two-leg ladder cuprates, a significant body of reliable and
reproducible data has been accumulated by using many probes, which
indicate that the doped two-leg ladder cuprates show many
nonconventional physical properties
\cite{dagotto1,dagotto2,katano,magishi,eccleston}. When charge
carriers are added to the two-leg ladder cuprates, a
metal-insulator transition is observed
\cite{uehara,nagata,isobe,piskunov,ohta,kato,nagata1}. Although
the ambient pressure ladder superconductivity was not observed
until now, superconductivity in one of the doped two-leg ladder
cuprate Sr$_{14-x}$Ca$_{x}$Cu$_{24}$O$_{41}$ has been observed
under pressure \cite{uehara,nagata,isobe,piskunov}. In particular,
the maximal superconducting (SC) transition temperature in
Sr$_{14-x}$Ca$_{x}$Cu$_{24}$O$_{41}$ occurs around the optimal
pressure, and then decreases in both underpressure and
overpressure regimes \cite{uehara,nagata,isobe,piskunov}.
Moreover, the structure of the doped two-leg ladder cuprates under
high pressure remains the same as the case in ambient pressure
\cite{isobe}, and the spin background in the SC phase does not
drastically alter its spin gap properties \cite{dagotto2}, which
show that the dominant physics arises from the individual ladders,
similar to the way that the dominant physics is determined by the
two-dimensional CuO$_{2}$ planes in the doped planar cuprate
superconductors \cite{kastner}. In this case, a challenging issue
for theory is to explain the pressure dependence of
superconductivity in the doped two-leg ladder cuprate
Sr$_{14-x}$Ca$_{x}$Cu$_{24}$O$_{41}$.

Experimentally, it has been shown
\cite{dagotto2,magishi,eccleston} that at ambient pressure, the
exchange coupling $J_{\parallel}$ along the legs is greater than
exchange coupling $J_{\perp}$ across a rung, i.e.,
$J_{\parallel}>J_{\perp}$, and similarly the hopping
$t_{\parallel}$ along the legs is greater than the rung hopping
strength $t_{\perp}$, i.e., $t_{\parallel}>t_{\perp}$. In this
case, the two-leg ladder cuprates are highly anisotropic
materials. Furthermore, the experimental results have showed that
the most important role of pressure for realizing
superconductivity in the doped two-leg ladder cuprates is to
reduce the distance between ladders and chains, and then the
coupling between ladders and chains is enhanced
\cite{uehara,nagata,isobe,piskunov,ohta,kato,nagata1}. This leads
to that the values of $J_{\perp}/J_{\parallel}$ and
$t_{\perp}/t_{\parallel}$ increase with increasing pressure. In
other words, the pressurization induces anisotropy shrinkage on
the two-leg ladder cuprates, and then there is a tendency toward
the isotropy for two-leg ladders
\cite{uehara,nagata,isobe,piskunov,ohta,kato,nagata1}. These
experimental results explicitly imply that the values of
$J_{\perp}/J_{\parallel}$ and $t_{\perp}/t_{\parallel}$ of the
doped two-leg ladder cuprates are closely related to the
pressurization, and therefore the pressure effects may be imitated
by the variation of the values of $J_{\perp}/J_{\parallel}$ and
$t_{\perp}/t_{\parallel}$.

Theoretically, it has been shown within the $t$-$J$ ladder model
that the charge carrier pair correlation is very robust
\cite{dagotto2,dagotto3}, clearly indicative of a ground-state
dominated by strong SC tendencies. In particular, it has been
found based on the density matrix renormalization group (DMRG) and
Monte Carlo simulations \cite{noack} that the pairing correlations
are enhanced when the top of the bonding quasiparticle band and
the bottom of the antibonding band are near the Fermi level.
Furthermore, using the renormalized mean-field theory, many
authors have shown that superconductivity should exist in the
d-wave channel \cite{sigrist}, which has been confirmed by variety
of numerical simulations \cite{riera}. However, to the best of our
knowledge, no systematic calculations have been performed within
the $t$-$J$ ladder model for the pressure dependence of the SC
transition temperature to confront the experimental data.

In this Letter, we study the pressure dependence of
superconductivity in the doped two-leg ladder cuprates within the
{\it anisotropic} $t$-$J$ ladder model. Our results show that the
SC transition temperature in the doped two-leg ladder cuprate
superconductors increases with increasing pressure in the
underpressure regime, and reaches a maximum in the optimal
pressure, then decreases in the overpressure regime, in
qualitative agreement with the experimental results \cite{isobe}.
Moreover, this domed shape of the pressure dependence of the SC
transition temperature is similar to that of the pressure
dependence of the longitudinal part of the SC gap parameter, and
therefore it is also shown that superconductivity in the doped
two-leg ladder cuprates is mainly produced by the development of
the pairing correlation along legs. The strong electron
correlation in the $t$-$J$ ladder model manifests itself by the
electron single occupancy local constraint
\cite{dagotto1,dagotto2}, this is why the crucial requirement is
to impose this electron local constraint for a proper
understanding of the physical properties of the doped two-leg
ladder cuprates. To incorporate this local constraint, we have
developed a charge-spin separation (CSS) fermion-spin theory
\cite{feng1}, where the constrained electron operators are
decoupled as the gauge invariant spinful fermion operator and spin
operator, with the spinful fermion operator keeps track of the
charge degree of freedom together with some effects of the spin
configuration rearrangements due to the presence of the doped hole
itself (dressed holon), while the spin operator keeps track of the
spin degree of freedom. The advantage of this approach is that the
electron local constraint for the single occupancy is satisfied in
analytical calculations \cite{feng1}. Within this CSS fermion-spin
theoretical framework, a kinetic energy driven SC mechanism
\cite{feng2} has been proposed, where the dressed holon-spin
interaction from the kinetic energy term induces the dressed holon
pairing state by exchanging spin excitations, then the electron
Cooper pairs originating from the dressed holon pairing state are
due to the charge-spin recombination, and their condensation
reveals the SC ground-state. Based on this kinetic energy driven
SC mechanism, we \cite{qin0} have discussed the possible doping
dependence of superconductivity in the doped two-leg ladder
cuprates within the {\it isotropic} $t$-$J$ ladder model, and the
result shows that the spin-liquid ground-state at the half-filling
evolves into the SC ground-state upon doping. Since a detailed
description of the method within the $t$-$J$ ladder model has been
given in Ref. \cite{qin0}, and therefore we do not repeat here.

\begin{figure}
\begin{center}
\includegraphics*[height=6.5cm,width=8cm]{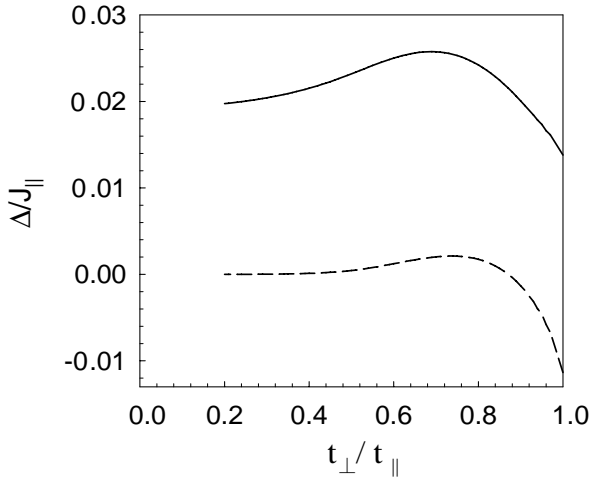}
\end{center}
\caption{The longitudinal (solid line) and transverse (dashed
line) SC gap parameters as a function of $t_{\perp}/t_{\parallel}$
at $\delta=0.25$ in $t_{\parallel}/J_{\parallel}=2.5$ with
$T=0.0001J_{\parallel}$.}
\end{figure}

In Fig. 1, we plot the longitudinal (solid line) and transverse
(dashed line) parts of the SC gap parameters as a function of
$t_{\perp}/t_{\parallel}$ at the doping concentration
$\delta=0.25$ in parameter $t_{\parallel}/J_{\parallel}=2.5$ with
temperature $T=0.0001J_{\parallel}$. For the convenience in the
discussions, we have chosen the variation of
$(t_{\perp}/t_{\parallel})^{2}$ under the pressure is the same as
that of $J_{\perp}/J_{\parallel}$, i.e.,
$(t_{\perp}/t_{\parallel})^{2}=J_{\perp}/J_{\parallel}$. Our
results show that both longitudinal and transverse parts of the SC
gap parameter have a similar pressure dependent behavior. In
particular, the value of the longitudinal part of the SC gap
parameter $\Delta_{L}$ increases with increasing
$t_{\perp}/t_{\parallel}$ in the lower $t_{\perp}/t_{\parallel}$
regime, and reaches a maximum in the optimal
$(t_{\perp}/t_{\parallel})_{{\rm opt}}\approx 0.7$, then decreases
in the higher $t_{\perp}/t_{\parallel}$ regime. In this case, the
underpressure, optimal pressure, and overpressure regimes are
corresponding to the lower $t_{\perp}/t_{\parallel}$, optimal
$(t_{\perp}/t_{\parallel})_{{\rm opt}}$, and higher
$t_{\perp}/t_{\parallel}$ regimes, respectively.

In correspondence with the SC gap parameter, the SC transition
temperature $T_{c}$ as a function of $t_{\perp}/t_{\parallel}$ at
$\delta=0.25$ in $t_{\parallel}/J_{\parallel}=2.5$ is plotted in
Fig. 2. For comparison, the experimental result \cite{isobe} of the
SC transition temperature $T_{c}$ in the doped two-leg ladder
cuprate superconductor Sr$_{14-x}$Ca$_{x}$Cu$_{24}$O$_{41}$ with
$x=13.6$ (the corresponding doping concentration $\delta\sim 0.25$)
as a function of pressure is also shown in Fig. 2 (inset). Our
result shows that the SC transition temperature in the doped two-leg
ladder cuprate superconductors increases with increasing
$t_{\perp}/t_{\parallel}$ in the lower $t_{\perp}/t_{\parallel}$
regime, and reaches a maximum in the optimal
$(t_{\perp}/t_{\parallel})_{{\rm opt}}\approx 0.7$, then decreases
in the higher $t_{\perp}/t_{\parallel}$ regime. Using a reasonably
estimative value of $J_{\parallel}\sim 90$meV$\approx 1000$K in the
doped two-leg ladder cuprate superconductors \cite{katano}, the SC
transition temperature in the optimal pressure is T$_{c}\approx
0.036J_{\parallel}\approx 36$K, in qualitative agreement with the
experimental data \cite{isobe}. Moreover, in comparison with the
results of the longitudinal and transverse SC gap parameters in Fig.
1, we therefore find that this domed shape of the pressure
dependence of the SC transition temperature is similar to that of
the pressure dependence of the longitudinal part of the SC gap
parameter. As the pressure increases, although the hole movement
along the rung direction is enhanced and superconductivity appears
in the ladder, our result also shows that superconductivity in the
doped two-leg ladder cuprates is mainly produced by the development
of the pairing correlation along legs, and is consistent with the
one-dimensional charge dynamics under high pressure
\cite{isobe,dagotto2}.

\begin{figure}
\begin{center}
\includegraphics*[height=6.5cm,width=8cm]{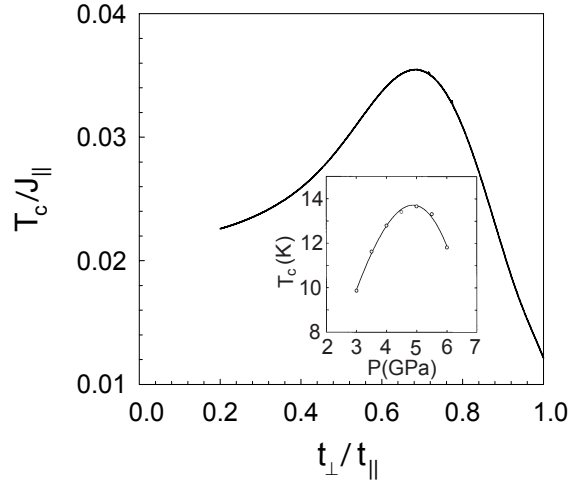}
\end{center}
\caption{The SC transition temperature as a function of
$t_{\perp}/t_{\parallel}$ at $\delta=0.25$ in
$t_{\parallel}/J_{\parallel}=2.5$. Inset: the experimental result
of Sr$_{14-x}$Ca$_{x}$Cu$_{24}$O$_{41}$ with $x=13.6$ (the
corresponding doping concentration $\delta\sim 0.25$) taken from
Ref. [8].}
\end{figure}

For a better understanding of superconductivity in the doped two-leg
ladder cuprates, we have made a series of calculations for the
doping dependence of the SC transition temperature at different
$t_{\perp}/t_{\parallel}$, and the result of the SC transition
temperature $T_{c}$ as a function of the hole doping concentration
$\delta$ for $t_{\parallel}/J_{\parallel}=2.5$ and
$(t_{\perp}/t_{\parallel})_{{\rm opt}}\approx 0.7$ is plotted in
Fig. 3. It is shown that the maximal SC transition temperature
T$_{c}$ occurs around the optimal doping concentration $\delta_{{\rm
opt}}\approx 0.12$, and then decreases in both underdoped and
overdoped regimes. Moreover, T$_{c}$ in the underdoped regime is
proportional to the hole doping concentration $\delta$, and
therefore T$_{c}$ in the underdoped regime is set by the hole doping
concentration as in the doped planar cuprates \cite{feng2}, which
reflects that the density of the dressed holons directly determines
the superfluid density in the underdoped regime. In comparison with
the result of the doped {\it isotropic} two-leg ladder case
\cite{qin0}, our present result also shows that the range of
superconductivity in the doped {\it anisotropic} $t$-$J$ ladder
cuprates is larger than that of the doped {\it isotropic} $t$-$J$
ladder case, in particular, the optimal doping in the present {\it
anisotropic} case moves to higher doping regime than that of the
doped {\it isotropic} $t$-$J$ ladder cuprates. Furthermore, this
domed shape of the doping dependence of the SC transition
temperature in Fig. 3 is similar to that of the pressure dependence
of the SC transition temperature in Fig. 2, reflecting a
corresponding relationship between the pressure effect and hole
doping concentration. This is consistent with the experiments
\cite{uehara,nagata,isobe,piskunov,ohta,kato,nagata1}, since it has
been shown from the experiments that the main effect of pressure in
the doped two-leg ladder cuprates is to reduce the distance between
the ladders and chains, which leads to the doped hole redistribution
between chains and ladders \cite{uehara,nagata,isobe,piskunov}, in
particular, when Ca is doped upon the original Sr-based Ca-undoped
phase, the interatomic distance ladder-chain was found to be reduced
by Ca substitution, leading to a redistribution of holes originally
present only on the chains \cite{ohta,kato}.

\begin{figure}
\begin{center}
\includegraphics*[height=6.5cm,width=8cm]{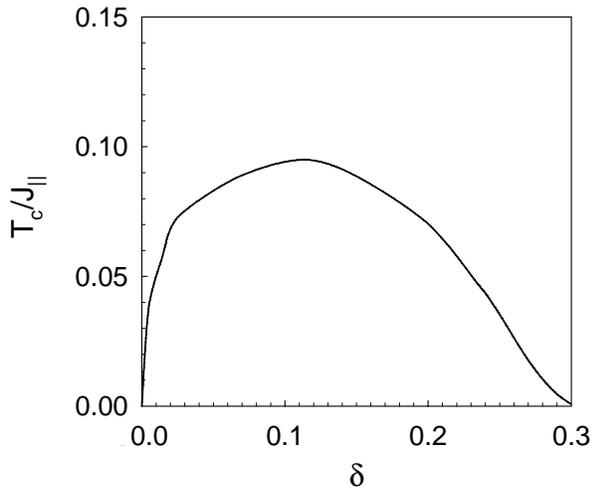}
\end{center}
\caption{The SC transition temperature as a function of the doping
concentration for $t_{\parallel}/J_{\parallel}=2.5$ and
$(t_{\perp}/t_{\parallel})_{{\rm opt}}\approx 0.7$. }
\end{figure}

As we \cite{qin0} have shown the SC-order in the doped two-leg
ladder cuprates is established through an emerging SC
quasiparticle, and therefore the SC-order is controlled by both
gap function and quasiparticle coherence. In this case, the
essential physics of the pressure dependence of the SC transition
temperature in the doped two-leg ladder cuprates can be understood
from a competition between the kinetic energy and magnetic energy
\cite{qin0}. As we have mentioned above, the experimental results
\cite{uehara,nagata,isobe,piskunov} have indicated that the main
effect of the pressure in the doped two-leg ladder cuprate
superconductors is to reduce the distance between the ladders and
chains (then enhance the coupling between the ladders and chains),
which leads to the doped hole redistribution between chains and
ladders. On the other hand, when Ca is doped upon the original
Sr-based Ca-undoped phase, the interatomic distance ladder-chain
was found to be reduced (then the coupling between the ladders and
chains is enhanced) by Ca substitution, leading to a
redistribution of holes originally present only on the chains
\cite{ohta,kato,nagata1}. These experimental results show that an
increase of the pressure (then an increase of the coupling between
the ladders and chains) may be corresponding to an increase in the
number of charge carriers on the ladders
\cite{isobe,piskunov,ohta,kato,nagata1}. In this case, the kinetic
energy increases with increasing pressure (doping), but at the
same time, the spin correlation is destroyed, therefore the
pressure effect (doping) on the doped two-leg ladder cuprates can
be considered as a competition between the kinetic energy and
magnetic energy, and the magnetic energy decreases with increasing
pressure (doping). In the underpressure (underdoping) and optimal
pressure (optimal doping) regimes, the charge carrier
concentration is small, and therefore magnetic energy is rather
large, then the dressed holon (then electron) attractive
interaction by exchanging spin excitations is also rather strong
to form the dressed holon pairs (then electron Cooper pairs) for
the most dressed holons (then electrons), therefore the SC
transition temperature increases with increasing pressure
(doping). However, in the overpressure (overdoping) regime, the
charge carrier concentration is large and magnetic energy is
relatively small, then the dressed holon (then electron)
attractive interaction by exchanging spin excitations is also
relatively weak, in this case, not all dressed holons (then
electrons) can be bounden as dressed holon pairs (then electron
Cooper pairs) by this weak attractive interaction, and therefore
the SC transition temperature decreases with increasing pressure
(doping).

Based on the DMRG and Monte Carlo numerical simulation
\cite{noack}, the domed shape of the pressure dependence of the SC
pairing correlation in the doped two-leg ladder cuprates has been
discussed within the Hubbard ladder model. They \cite{noack}
calculate the ground state expectation value of the rung-rung
pair-field correlation function $D(i,j)=\langle
\Delta(i)\Delta^{\dagger}(j)\rangle$, and show that near the
half-filling, the strength of the pairing correlation depends
sensitively upon $t_{\perp}/t_{\parallel}$ and the doping as well
as $U/t_{\parallel}$ (then $t_{\parallel}/J_{\parallel}$), where
there are strong interchain antiferromagnetic correlations and a
large single-particle spectral weight at the Fermi points of the
bonding and antibonding bands, which is consistent with our
present result. However, their result also shows \cite{noack} that
the optimal value of the SC pairing correlation occurs for
intermediate values of $U/t_{\parallel}$ and for doping near
half-filling with $t_{\perp}/t_{\parallel}=1.5$, which is
inconsistent with our present result. The reason for this
inconsistency is not clear, and the related issue is under
investigation now.

In summary, we have discussed the effect of the pressure on
superconductivity in the doped two-leg ladder cuprates within the
kinetic energy driven SC mechanism. Our results show that the SC
transition temperature in the doped two-leg ladder cuprates
increases with increasing pressure in the underpressure regime,
and reaches a maximum in the optimal pressure, then decreases in
the overpressure regime. This domed shape of the pressure
dependence of the SC transition temperature is similar to that of
the pressure dependence of the longitudinal part of the SC gap
parameter, and therefore it is also shown that the pressure
dependence of superconductivity in the doped two-leg ladder
cuprates is mainly produced by the development of the pairing
correlation along legs.

\begin{ack}
The authors would like to thank Dr. Huaiming Guo and Dr. Yu Lan
for the helpful discussions. This work was supported by the
National Natural Science Foundation of China under Grant Nos.
10547104 and 90403005, and the funds from the Ministry of Science
and Technology of China under Grant No. 2006CB601002.
\end{ack}

\end{document}